\documentclass[a4paper,fleqn,usenatbib,useAMS]{mnras}

\usepackage{graphicx}	
\usepackage{amsmath}	
\usepackage{amssymb}	

\usepackage[T1]{fontenc}
\usepackage{ae,aecompl}

\title[$\gamma$-Ray Pulse Profile of the Crab pulsar with CTA]{Simulated Gamma-Ray Pulse Profile of the Crab pulsar with the Cherenkov Telescope Array}

\author[A. Burtovoi and L. Zampieri]{A. Burtovoi$^{1,2}$\thanks{E-mail: \href{mailto:aleksandr.burtovoi@studenti.unipd.it}{aleksandr.burtovoi@studenti.unipd.it}} and L. Zampieri$^{2}$\thanks{E-mail: \href{mailto:luca.zampieri@oapd.inaf.it}{luca.zampieri@oapd.inaf.it}}
\\
$^{1}$Department of Physics and Astronomy, University of Padova, vicolo dell' Osservatorio 3, I-35122 Padova, Italy\\
$^{2}$INAF - Astronomical Observatory of Padova, vicolo dell' Osservatorio 5, I-35122 Padova, Italy
}

\pubyear{2016}

\begin{document}
\label{firstpage}
\pagerange{\pageref{firstpage}--\pageref{lastpage}}
\maketitle

\begin{abstract}
We present simulations of the very high energy (VHE) gamma-ray light curve of the Crab pulsar as observed by the Cherenkov Telescope Array (CTA). The CTA pulse profile of the Crab pulsar is simulated with the specific goal of determining the accuracy of the position of the interpulse. We fit the pulse shape obtained by the MAGIC telescope with a three-Gaussian template and rescale it to account for the different CTA instrumental and observational configurations. Simulations are performed for different configurations of CTA and for the ASTRI mini-array. The northern CTA configuration will provide an improvement of a factor of $\sim$3 in accuracy with an observing time comparable to that of MAGIC (73 hours). Unless the VHE spectrum above 1 TeV behaves differently from what we presently know, unreasonably long observing times are required for a significant detection of the pulsations of the Crab pulsar with the high-energy-range sub-arrays. We also found that an independent VHE timing analysis is feasible with Large Size Telescopes (LSTs). CTA will provide a significant improvement in determining the VHE pulse shape parameters necessary to constrain theoretical models of the gamma-ray emission of the Crab pulsar. One of such parameters is the shift in phase between peaks in the pulse profile at VHE and in other energy bands that, if detected, may point to different locations of the emission regions. 
\end{abstract}

\begin{keywords}
pulsars: individual: Crab pulsar - gamma rays: stars
\end{keywords}

\begingroup
\let\clearpage\relax
\endgroup
\newpage

\section{Introduction}\label{sec:1}
The Crab pulsar (PSR J0534+2200) was the first pulsar to be detected by Cherenkov telescopes at very high energy (VHE) gamma-rays above a few tens of GeV \citep[e.g.][]{Aliu2008,Aleksic2011,Aliu2011,Aleksic2012}. It is the compact remnant of a supernova which exploded approximately 1000 years ago at a distance of about 2 kpc from the Sun. The magnetic field of the pulsar is $3.8 \times 10^{12}$ G, its rotational period $\sim$33.62 ms, and its spin-down power $\sim$$4.6 \times 10^{38}~\mathrm{erg~s}^{-1}$ \citep{Manchester2005}. The Crab pulsar is detected at all wavelengths from radio to TeV gamma-rays.

Investigating the pulse profile of pulsars in different energy bands is important in order to fully understand the physical mechanisms responsible for accelerating particles to relativistic energies. Several groups have studied the pulse profile of the Crab pulsar at different energies. The timing properties in the radio band were investigated with a number of radio telescopes, including the Nan\c{c}ay French telescope \citep{Theureau2005} and the Jodrell Bank Observatory \citep{Hobbs2004}. Some of the most accurate optical observations of the Crab pulsar, with time resolutions of hundreds of picoseconds, were recently carried out with the Copernico Telescope in Asiago \citep{Germana2012} and the New Technology Telescope in La Silla \citep{Zampieri2014}. Detailed X-ray pulse profiles were obtained with the Rossi X-ray Timing Explorer (RXTE) \citep{Rots2004} as well as with \textit{Suzaku} \citep{Terada2008}, \textit{Swift} \citep{Cusumano2012} and \textit{XMM-Newton} \citep{Kirsch2006}. Hard X-ray (100--200 keV) and soft gamma-ray (0.75--30 MeV) observations were carried out with INTEGRAL \citep{Mineo2006} and COMPTEL \citep{Kuiper2001}, respectively. Gamma-ray ($>$100 MeV) observations were performed by AGILE \citep{Pellizzoni2009} and \textit{Fermi}-LAT \citep{Abdo2010}. Finally, recent observations with ground based Cherenkov telescopes, such as MAGIC \citep{Aleksic2012,Aleksic2014_2,Ahnen2015arx} and VERITAS \citep{Aliu2011}, have obtained pulse profiles of the Crab pulsar in VHE gamma rays.

The gamma-ray spectrum of the Crab pulsar above 10 GeV is not consistent with the exponential or steeper cut-off inferred from \textit{Fermi}-LAT data in the 100 MeV -- 100 GeV energy range \citep{Abdo2010}. MAGIC and VERITAS observations show that the amplitude of the main pulse of the Crab pulsar is lower than the amplitude of the interpulse contrary to what is observed at lower energies with \textit{Fermi}-LAT.

Although there is still no comprehensive theory that can describe the overall emission properties of the Crab pulsar, VHE observations constrain significantly the models of pulsar emission \citep[e.g.][]{Aharonian2012, Lyutikov2012}.

The Cherenkov Telescope Array (CTA), currently in the development stage, is a project which aims at building two arrays, one in each hemisphere, of imaging atmospheric Cherenkov telescopes (IACTs, \citealt{CTA2011}). This observatory is designed to improve the capabilities of present Cherenkov imaging telescopes (MAGIC, VERITAS, H.E.S.S.) and will allow us to explore VHE gamma-ray phenomena in more detail. CTA (North+South) will comprise $\sim$140 telescopes of three different types (Large, Medium and Small Size Telescopes\footnote{SSTs are expected to be deployed only at the southern site.} with diameters of 23, $\sim$10-12 and 4 meters, respectively). This will allow CTA to cover the full sky over the energy range from a few tens of GeV to more than 100 TeV \citep{Acharya2013, Bernlohr2013} and to reach 10 times better sensitivity and angular resolution compared to present Cherenkov telescopes installations. There is a possibility of dividing the whole array into different sub-arrays corresponding to different mirror sizes: the LST-array, the MST-array and the SST-array consisting of only Large, Medium and Small Size Telescopes, respectively.

As part of the CTA project, a dual-mirror prototype of the Small Size Telescope is under development within the framework of the ASTRI (\textit{Astrofisica con Specchi a Tecnologia Replicante Italiana}) flagship project of the Italian Ministry of Research and Education led by INAF \citep{LaPalombara2014}. This project foresees the construction of a mini-array of 9 telescopes, the ASTRI mini-array \citep{Vercellone2015arx}, possibly as a first segment of the southern CTA installation.

The first comprehensive investigation of  prospects for VHE observations of pulsars (including the Crab pulsar) is reported in \citet{Ona2013}. The unprecedented sensitivity achievable with CTA prompted us to perform a quantitative investigation of the pulse shape and timing of the Crab pulsar at VHE attainable with the CTA observatory. To estimate the impact of CTA, various simulations were performed for different array configurations and exposure times.

The outline of the paper is as follows. In Sect. \ref{sec:2} we present the algorithm used to simulate Crab pulsar observations with CTA. A short description of the different CTA configurations is given in Sect. \ref{sec:3}. The pulse profiles resulting from the simulations are presented in Sect. \ref{sec:4}. In Sect. \ref{sec:5} we discuss our results, while conclusions follow in Sect. \ref{sec:6}.

\section{VHE pulse profile of the Crab pulsar}\label{sec:2}
\begin{figure}
	\centering
	\includegraphics[width=0.50\textwidth]{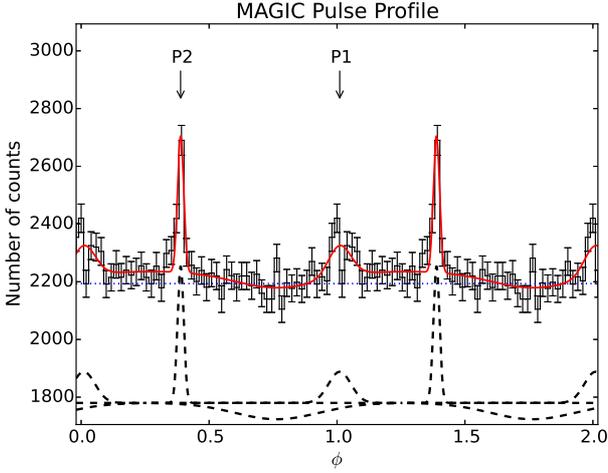}
	\caption{MAGIC 50--400 GeV pulse profile of the Crab pulsar \citep{Aleksic2012}, along with the fitting function $I$ (red solid line; see text for details). The black dashed lines represent the Gaussian components of $I$, while the blue dotted line is the background level. P1 and P2 are the main pulse and interpulse, respectively. The adopted number of bins per period is 51. \textit{(A color version of this figure is available on-line)}.} 
   	\label{fig:Interp_M}
\end{figure}

The Crab pulsar region has been observed with several VHE telescopes (H.E.S.S., \citep{Aharonian2006_2,Abramowski2014}; HEGRA, \citep{Aharonian2004}; Whipple, \citep{Weekes1989,Grube2008}; CAT, \citep{Masterson2001}; MAGIC, \citep{Aleksic2012,Aleksic2014_2,Aleksic2015}; VERITAS, \citep{Aliu2011}). For the sake of comparison, in the following we will consider as reference the observations carried out with the two MAGIC telescopes located in La Palma during the period between the winter season 2009/2010 and that of 2010/2011 \citep{Aleksic2012}. The energy range is 50--400 GeV. A light curve was obtained by phase folding approximately 73 hours of observations and is shown in Fig. \ref{fig:Interp_M}. The light curve can be quite reasonably reproduced by the sum of Gaussian functions plus a constant. Three Gaussians are sufficient for an accurate fit: the first two components correspond to the pulsar peaks (P1 and P2), while the third one (with negative amplitude) improves the fit in the off-pulse interval between 0.52 and 0.87. The adopted fitting function written as a function of phase $\phi$ is:
\begin{equation}
	I (\phi) = \sum_{i=1}^3 k_i G_{m_i, s_i} (\phi) + C\, ,
	\label{eq:1}
\end{equation}
where $G_{m_i, s_i} (\phi) = 1/(\sqrt{2\pi}s_i) \exp[- (\phi - m_i)^2/(2s_i^2)]$ is a Gaussian function with mean $m_i$, standard deviation $s_i$, and normalization $k_i$ (Table \ref{tab:Inter_Params}), while the constant $C$ accounts for the background. Taking $m_i$, $s_i$, $k_i$ and $C$ as free parameters, the MAGIC pulse profile is well fitted by Eq. (\ref{eq:1}) with a reduced $\chi^2$ of 1.07. Hereafter we fix the values of the parameters obtained from the fit and use them in the simulations of pulse profiles of the Crab pulsar as would be observed with CTA.

\begin{table}
\caption{Parameters of the fitting function $I$, given by the sum of three Gaussians and a constant. The mean, standard deviation and normalization of the Gaussians are reported in the first, second and third column, respectively. The value of $C$ is listed in the forth row.}
\label{tab:Inter_Params}
\centering
\begin{tabular}{c c c c}
\hline\hline
$i$ & $m_i$ & $s_i$ & $k_i$ \\
\hline
1 & $0.389\pm0.002$	& $0.012\pm0.002$& $14\pm2$	\\
2 & $1.01\pm0.02$		& $0.04\pm0.02$ 	& $12\pm7$ 	\\
3 & $0.76\pm0.09$ 	& $0.16\pm0.13$ 	& $-23\pm22$	\\
\hline
\multicolumn{4}{l}{$C=(2.24\pm0.01)\times10^3$}\\
\hline
\end{tabular}
\end{table}

To simulate the pulse profile observed by a CTA-like instrument we calculate the background level $I^{\rm M}_{\rm BG}$ of the MAGIC data in the off-pulse region from phase 0.52 to 0.87 and subtract it from the fitted profile $I$. Then, we rescale the profile $I$ according to the different effective area $A_{\rm eff}$ of the CTA configurations and to the different observation durations $t_{\rm obs}$. We also assume that the pulse shape does not strongly depend on energy between 0.01 TeV and $\sim$100 TeV, which allows us to rescale the profile with the number of counts in different spectral bands. Although there is some evidence of evolution of the pulse shape in gamma rays (e.g. \citealt{Aleksic2014_2}), this appears to be in the direction of increasing the significance of the interpulse (with respect to the main pulse). Thus, the actual detection of the interpulse with CTA, on which we will focus below, may in fact be more significant, and our estimates may then be regarded as conservative. To determine the rescaling factor we need an estimate of the energy spectrum in the CTA energy range, which is the major source of uncertainty in the present calculation.

Assuming that $F(E)$ is an appropriate representation of the actual phase-averaged pulsar spectrum in the CTA energy range, we can then rescale the pulse shape according to the following expression: 
\begin{equation}
	I^{'} = I \times  \frac {\int_{E_{\mathrm{min}}}^{E_{\mathrm{max}}} F(E)\,A_{\rm eff}(E)\,t_{\rm obs}\, \mathrm{d}E}{\int_{E_{\mathrm{min}}^{\mathrm{M}}}^{E_{\mathrm{max}}^{\mathrm{M}}} F_{\rm M}(E)\, A_{\rm eff}^{\rm M}(E)\,t_{\rm obs}^{\rm M}\, \mathrm{d}E},
	\label{eq:3}
\end{equation}
where $F(E)$ ($F_{\rm M}(E)$) is the CTA (MAGIC) spectrum, $E_{\mathrm{min}}$ ($E_{\mathrm{min}}^{\mathrm{M}}$) and $E_{\mathrm{max}}$ ($E_{\mathrm{max}}^{\mathrm{M}}$) bracket the corresponding energy range, $A_{\rm eff}$ ($A_{\rm eff}^{\rm M}$) and $t_{\rm obs}$ ($t_{\rm obs}^{\rm M}=72.78$ hours) are the effective area and corresponding observing time in the CTA (MAGIC) configuration. The rescaling factor (the ratio of the two integrals) is determined by comparing the number of counts of the simulated configuration with that of MAGIC. The calculation is done adopting effective areas for similar zenith angles (20$^\circ$ for CTA and VERITAS, averaged below 30$^\circ$ for MAGIC).

For $F(E)$ we assume a power law:
\begin{equation}
	F(E) = \frac{\mathrm{d}N}{\mathrm{d}E} = N_0 \times \left( \frac{E}{0.1~\mathrm{TeV}} \right)^{-\Gamma},
	\label{eq:69.31}
\end{equation}
where $N_0$ and $\Gamma$ are the normalization and spectral index, respectively. The values of $N_0$ and $\Gamma$ are taken from \citet{Aleksic2012}: $N_0 = (13.0 \pm 1.6) \times10^{-11}$ TeV$^{-1}$ cm$^{-2}$ s$^{-1}$ and $\Gamma = 3.57 \pm 0.27$\footnote{Only statistical errors are quoted.}. We do not use values from the recent work by \citet{Ahnen2015arx} because we need spectral parameters averaged over the emission of the two peaks, while they analyzed the main pulse and interpulse separately.

Another parameter required to estimate the actual light curve observed with CTA is the background emission, which is generally dominated by the Crab Nebula rather than by backgrounds particles (hadrons, electrons and diffuse gamma rays). We determine it by adopting a simplified approach, similar to that outlined above for rescaling the source counts since, at CTA resolution, both the pulsar and surrounding nebula can be considered as point-sources. Assuming that the VHE emission of the Crab Nebula dominates over cosmic ray background up to the $\sim$100 TeV, the background is obtained by re-normalizing the counts of the Nebula spectrum in the different energy ranges (similar to Eq. (\ref{eq:3})):
\begin{equation}
	I_{\rm BG} = I_{\rm BG}^{\mathrm{M}}\, \times  \frac {\int_{E_{\mathrm{min}}}^{E_{\mathrm{max}}} F_{\rm BG}(E)\,A_{\rm eff}(E)\,t_{\rm obs}\,  \mathrm{d}E}{\int_{E_{\mathrm{min}}^{\mathrm{M}}}^{E_{\mathrm{max}}^{\mathrm{M}}}  F_{\rm BG}(E)\,A_{\rm eff}^{\rm M}(E)\,t_{\rm obs}^{\rm M}\,  \mathrm{d}E},
	\label{eq:6}
\end{equation}
where $I^{\rm M}_{\rm BG}$ is the MAGIC background, measured in the off-pulse region from phase 0.52 to 0.87 (blue dotted line in Fig. \ref{fig:Interp_M}; \citealt{Aleksic2012}). For the nebular spectrum $F_{\rm BG}(E)$, we take the log-parabola approximation of \citet{Aleksic2015}:
\begin{eqnarray}
	F_{\rm BG}(E) = && (3.23\pm0.03)\times 10^{-11}\,   \times \nonumber \\
                                        && \left( \frac {E}{1 \, {\rm TeV}} \right)^{
                                           -(2.47\pm0.01)-(0.24\pm0.01)\log(E/1 \, {\rm TeV})} \nonumber \\
                                        && {\rm TeV}^{-1} {\rm cm}^{-2} {\rm s}^{-1} \,  .
	\label{eq:5.1}
\end{eqnarray}

\begin{figure}
	\centering
	\includegraphics[width=0.50\textwidth]{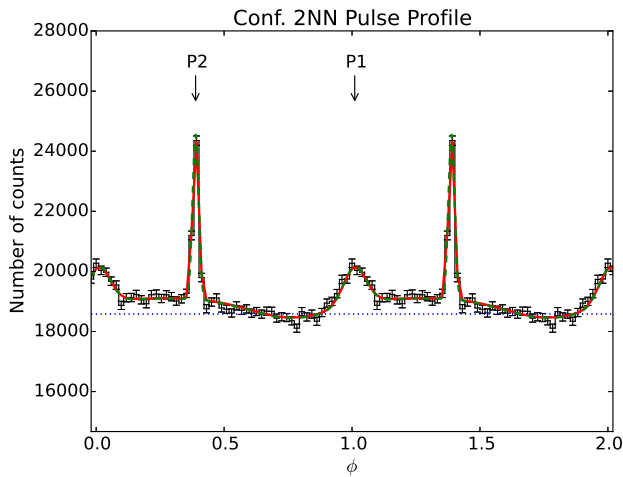}
	\caption{Simulated ($I_s$, black histogram) and assumed CTA pulse profile ($I^{'}$, red line) of the Crab pulsar detected by Conf. 2NN during an observation with duration $t_{\rm obs}^{\rm M} = 72.78$ hours and using 51 bins per period. The green dashed line is a best fit with three Gaussians of the simulated pulse profile. The blue dotted line is the background level. P1 and P2 represent the main pulse and interpulse, respectively. \textit{(A color version of this figure is available on-line)}.}
	\label{fig:Sim_PulsePrifile}
\end{figure}

Summarizing, we generate the simulated pulse profile $I_s$ detected by CTA using the following procedure:
\begin{enumerate}
	\item We approximate the pulse profile of the Crab pulsar with the fitting function $I$ (Eq. (\ref{eq:1}) and Fig. \ref{fig:Interp_M}).
	\item We then calculate the MAGIC background level $I^{\rm M}_{\rm BG}$ in the off-pulse region from phase 0.52 to 0.87 and subtract it from the pulse profile $I$. With this value we then computed the rescaled pulse shape $I^{'}$ from Eq. (\ref{eq:3}).
	\item We calculate the CTA background level $I_{\rm BG}$ from Eq. (\ref{eq:6}) and add it to the pulse profile $I^{'}$.
	\item Stochastic properties are added to the pulse shape $I^{'}$ to produce the final simulated signal $I_s$. The simulated pulse profile in the $i$-th bin, $I_{s,i}$, is considered to be a random value following a Gaussian distribution with mean value equal to $I^{'}$ and standard deviation equal to $\sqrt{I^{'}}$. The error in each bin is the square root of the number of counts in that bin, $\sqrt{I_{s,i}}$. An example of the final simulated pulse profile is shown in Fig. \ref{fig:Sim_PulsePrifile}.
\end{enumerate}

\section{CTA and ASTRI mini-array configurations}\label{sec:3}
Simulated pulse profiles of the Crab pulsar were computed for different sub-arrays of CTA. The corresponding configurations and their properties are listed in Table \ref{tab:1}. We consider standard arrays and sub-arrays, which consist of telescopes of different sizes (LSTs and MSTs) distributed according to Monte-Carlo Prod2 configuration (Conf.) 2NN, representative of the northern CTA installation (from Leoncito++ package\footnote{Monte-Carlo Prod2 DESY (Sep 2014) package available at \url{http://www.cta-observatory.org/ctawpcwiki/index.php/WP_MC\#Interface_to_WP_PHYS}.}). We also consider sub-arrays with identical types of telescopes, such as Large Size Telescopes (2NN-LST or LST-array) and Medium Size Telescopes (2NN-MST or MST-array). In addition, Conf. 2e -- a possible configuration of CTA-South -- is of great interest and is then included for comparison\footnote{We note that calculations for this configuration are performed for a fixed zenith angle (20$^\circ$). Although for CTA-South this is smaller than the actual zenith angle of the Crab pulsar, such simulations are performed only for comparison purposes and are not meant to provide detailed quantitative assessments of the pulse profile, detected at the southern installation.}. Finally, we use a distribution of 9 identical 4-meter SSTs with a separation of 257 meters (Conf. s9-4-257m) as an appropriate representation of the ASTRI mini-array in the present MC-Prod2 package.

\begin{table}
\caption{Configurations of MAGIC, VERITAS and different sub-arrays of CTA simulated in MC-Prod2.}
\label{tab:1}
\centering
\begin{tabular}{l c c c c}
\hline\hline
Name 		& Telescopes 		& Energy range & $E_{\rm thr}$	& $\left< A_{\rm eff} \right >_{\rm sp}$	\\
			&				& (TeV) 			& (TeV)			& [$10^5 \times \text{m}^2$]	\\
\hline
MAGIC 		& $2\times17$ m  	& 0.05--0.4	& 0.072	& 0.07	\\
VERITAS		& $4\times12$ m 	& 0.1--0.4		& 0.136	& 0.28	\\
LST-array		& 4 LST 			& 0.04--158	& 0.040	& 0.49	\\
MST-array	& 14 MST			& 0.1--158	& 0.158	& 0.71	\\
Mini-array	& 9 SST	 		& 1.6--158	& 3.981	& 0.71	\\
&&& \\
Conf. 2NN	& 4 LST 			& 0.04--158	& 0.040	& 0.53	\\
			& 14 MST 		&			& 		&		\\
&&& \\
Conf. 2e		& 4 LST 			& 0.04--158	& 0.040	& 0.50	\\
			& 24 MST			&			& 		&		\\
			& 72 SST			&			& 		&		\\
\hline
\multicolumn{5}{p{0.45\textwidth}}{\textbf{Notes.} Configurations 2NN, LST-array and MST-array refer to the northern CTA installation. Conf. 2e corresponds to CTA-South. These configurations are taken from the MC-Prod2 DESY simulation package (\url{http://www.cta-observatory.org/ctawpcwiki/index.php/WP_MC\#Interface_to_WP_PHYS}). LST: Large Size Telescope with diameter 23 m. MST: Medium Size Telescope with diameter 12 m. SST: Small Size Telescope with diameter 4 m. As the best representation for the ASTRI mini-array (Mini-array), we consider a configuration of 9 SSTs from the same MC-Prod2 simulations (Conf. s9-4-257m). The energy ranges for all these configurations are taken from the corresponding instrument response functions, while those of MAGIC and VERITAS correspond to the energies at which the Crab pulsar spectrum was measured (see \citet{Aleksic2012} and \citet{Aliu2011}, respectively). $E_{\rm thr}$ is the energy threshold, while $\left< A_{\rm eff} \right >_{\rm sp}$ is the spectrum-weighted effective area of each configuration.}
\end{tabular}
\end{table}

The effective areas $A_{\rm eff}(E)$ and energy ranges ($E_{\rm min}$, $E_{\rm max}$) needed for the convolution with the source and background spectra (Eqs. (\ref{eq:3}) and (\ref{eq:6})) are inferred from the instrument response functions (IRFs) and are reported in Table \ref{tab:1}. They are calculated from simulations of 50-hour observations of a source emitting 1 Crab Unit\footnote{1 Crab Unit = $2.79\times10^{-11} \times (E/\text{1 TeV})^{-2.57}$ cm$^{-2}$ s$^{-1}$ TeV$^{-1}$.} at a 20 degree zenith angle and with a sensitivity averaged over north and south pointings. For the effective areas of MAGIC ($A_{\rm eff}^{\rm M}(E)$) and VERITAS ($A_{\rm eff}^{\rm V}(E)$) we adopt published values from \citet{Aleksic2012_2} and \citet{Kieda2013arx}, while for the corresponding energy ranges, in which the Crab pulsar spectra were measured, we refer to \citet{Aleksic2012} and \citet{Aliu2011}.

In Table \ref{tab:1} we report also the threshold energy $E_{\rm thr}$ and the spectrum-weighted effective area $\left< A_{\rm eff} \right >_{\rm sp}$ of each configuration. The former is the energy at which the product of the effective area with the source spectrum $F(E)$ (defined in Sect. \ref{sec:2}) peaks, while the latter is defined as:
\begin{equation}
	\left< A_{\rm eff} \right>_{\rm sp} = \frac{\int_{E_{\rm min}}^{E_{\rm max}} A_{\rm eff}(E) F(E) \mathrm{d}E}{\int_{E_{\rm min}}^{E_{\rm max}} F(E) \mathrm{d}E} \, .
	\label{eq:5.2}
\end{equation}
The values of $E_{\rm thr}$ for MAGIC and VERITAS are consistent with the corresponding values reported in \citet{Aleksic2012_2} and \citet{Aliu2011}, respectively.

\section{Results}\label{sec:4}
Simulated pulse profiles for each CTA array configuration are computed as described in Sect. \ref{sec:2} and are then fitted with the model function in Eq. (\ref{eq:1}). An example of such a calculation is shown in Fig. \ref{fig:Sim_PulsePrifile} for CTA Conf. 2NN.

An important quantity to constrain the parameters (e.g. the height and location of the emission region) of pulsar models is the difference in the time of arrival of the peaks in different energy bands \citep[e.g.][]{Oosterbroek2008, Aharonian2012}. At optical wavelengths \citet{Shearer2003} and \citet{Hinton2006} found a radio delay between the time of arrival of the optical and radio peaks of the order of 100 $\mu$s with an uncertainty of a few tens of microseconds, whereas no delay ($-60\pm50~\mu$s) was found by \citet{Golden2000}. While a secular change of this delay may be possible, within measurement uncertainties present observations appear to give delays consistently of the order of $\sim$150-250 $\mu$s, with the optical leading the radio \citep[e.g.][]{Oosterbroek2008, Zampieri2014}.

\begin{figure}
	\centering
	\includegraphics[width=0.5\textwidth]{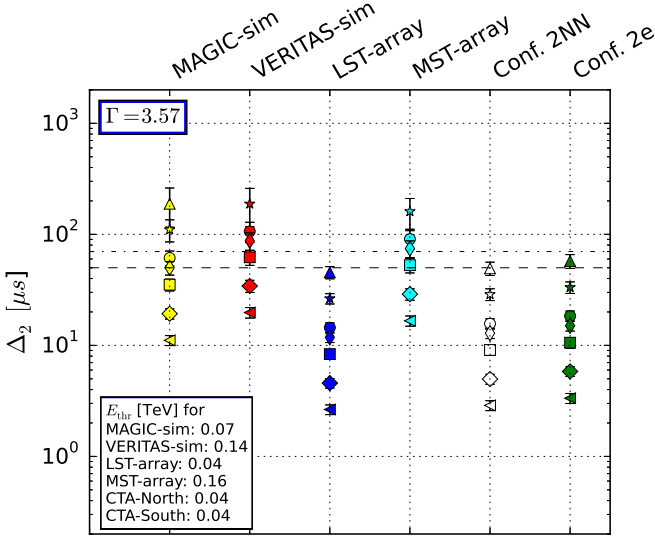}
	\caption{Uncertainty in the position of the interpulse P2 ($\Delta_2$) of the VHE profile of the Crab pulsar, simulated for MC-Prod2 Conf. 2NN (white) and for Conf. 2e (green). Results for MAGIC (yellow), VERITAS (red) and the CTA LST-array (blue) and MST-array are also shown. The spectral index of the Crab pulsar spectrum used in these simulations is $\Gamma$=3.57 (see Sect. \ref{sec:2}). Different markers correspond to observations of different durations in units of the MAGIC observing time ($t_{\rm obs}^{\rm M} = 72.78$ hours): 0.1 (triangles), 0.3 (stars), 1 (circles), 1.5 (thin diamonds), 3 (squares), 10 (diamonds), 30 (rotated triangles). Error-bars represent the standard deviation calculated from a set of simulations. The dashed and dotted-dashed lines show the uncertainties of the MAGIC ($\Delta_2 = 50~\mu$s; \citealt{Aleksic2012}) and VERITAS ($\Delta_2 =70~\mu$s; \citealt{Aliu2011}) observations, respectively.  \textit{(A color version of this figure is available on-line).}}
	\label{fig:Final_P2_A}
\end{figure}

Measuring an accurate time differences at VHE requires not only a precise time stamp (in CTA it will be of the order of ns), but also a good signal to noise ratio, or in other words good sensitivity. To check how well CTA can measure the peak positions with different observing times, we determined the position of the interpulse P2 and its error ($\Delta_2$) for the CTA sub-arrays considered here. Results are shown in Fig. \ref{fig:Final_P2_A}. The value of the uncertainty $\Delta_2$ clearly affects the accuracy with which it will be possible to perform this type of measurement.

For each configuration we repeated the simulations several times and then computed the average $\Delta_2$ and its statistical uncertainty. As can be seen from Fig. \ref{fig:Final_P2_A}, the accuracy of the pulse shape attainable by the CTA-North Conf. 2NN is such that even rather short observations (several hours; triangles) are sufficient to measure the position of the interpulse rather accurately. As a consistency check, we also simulated the MAGIC and VERITAS pulse profile. The simulated 73-hour MAGIC observations give results comparable to those of the real 73-hour observation ($\Delta_2=60 \pm 9~\mu$s versus $\sim$50 $\mu$s). Similar conclusions are reached for the simulations of the 110-hour VERITAS observations ($\Delta_2=90\pm20~\mu$s versus $\sim$70 $\mu$s).

We found that observations of short duration with VERITAS ($0.1\times t_{\rm obs}^{\rm M}$), the MST-array ($0.1\times t_{\rm obs}^{\rm M}$) are not sufficient to detect significant pulsations. Indeed, in these cases the energy threshold is higher than that of MAGIC. Simulations performed for the high-energy-range arrays containing only SSTs (e.g. ASTRI mini-array) with $E_{\rm thr}>1$ TeV yield no detection of pulsations even for very long observing times ($30\times t_{\rm obs}^{\rm M}$) and, therefore, are not shown in Fig. \ref{fig:Final_P2_A}. For the ASTRI mini-array significant pulsations are detected only for unrealistically long observing times of more than $\sim$$10^6$ hours.

In addition to the array configurations listed above, we repeated the simulations for different energy ranges of Conf. 2NN and also for other CTA configurations, such as Confs. 2Nc, 2Ne, 2Nb, 2Nd, 2Nf -- representatives of the northern CTA installation --, and Confs. 2b, 2c -- possible layouts of CTA-South. Results are reported in Appendix \ref{app:1}.

We investigated the possibility of measuring an energy dependent shift in the position of the interpulse with different CTA sub-arrays. In particular, we study whether it would be possible to measure phase shifts in the pulse profile among arrays made entirely by different types of telescopes (LSTs, MSTs, SSTs), which are most sensitive in the different energy ranges. This measurement is feasible in $\sim$73 hours with the LSTs and MSTs, but not with the SSTs (because of the larger amount of time required for such array to detect pulsations, see Sect. \ref{sec:5} for details). The uncertainty in measuring the position of the peak of the pulse profile with the LST- and MST- arrays is $\Delta_2\sim14~\mu$s and $\Delta_2\sim90~\mu$s, respectively (blue and cyan circles in Fig. \ref{fig:Final_P2_A}). Therefore, the error on the measurement of the shift between the time of arrival of the LST and MST interpulses is about $\sqrt{14^2 + 90^2} \approx 90~\mu$s. Any potential phase shift larger than $3\times90~\mu$s (=270 $\mu$s) between the pulses measured at $\sim$40 GeV with the LSTs and at $\sim$100 GeV with the MSTs would be measurable in 73 hours at the 3$\sigma$ confidence level.

\begin{figure}
	\centering
	\includegraphics[width=0.5\textwidth]{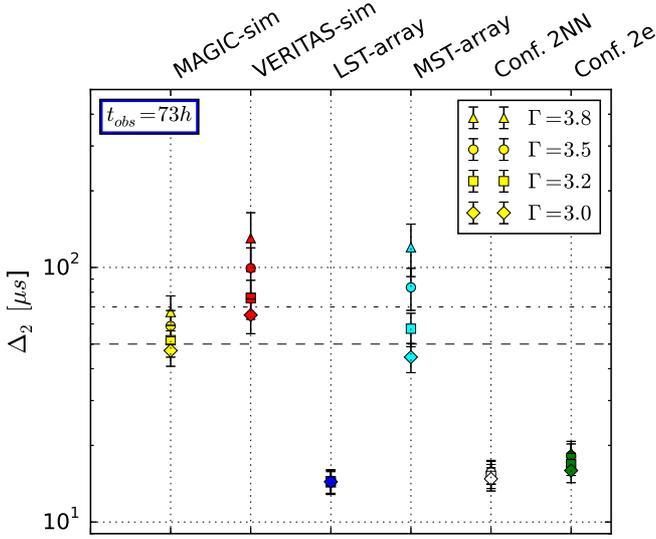} 
	\caption{Uncertainty in the position of the interpulse P2 ($\Delta_2$) of the VHE profile of the Crab pulsar for different values of the spectral index $\Gamma$: 3.8 (triangles), 3.5 (circles), 3.2 (squares), 3.0 (diamonds). Results for MAGIC (yellow), VERITAS (red), the LST-array (blue), the MST-array (cyan), Conf. 2NN (white) and Conf. 2e (green) are shown. Simulations are performed in the full energy range of each configuration. Observing time is $t_{\rm obs} = 73$ hours. Error-bars represent the standard deviation calculated from a set of simulations. The dashed and dotted-dashed lines show the uncertainties of the MAGIC ($\Delta_2 = 50~\mu$s; \citealt{Aleksic2012}) and VERITAS ($\Delta_2 = 70~\mu$s; \citealt{Aliu2011}) observations, respectively.  \textit{(A color version of this figure is available on-line).}}
	\label{fig:D2_ind_Prod2_Array_New}
\end{figure}

\begin{figure}
	\centering
	\includegraphics[width=0.5\textwidth]{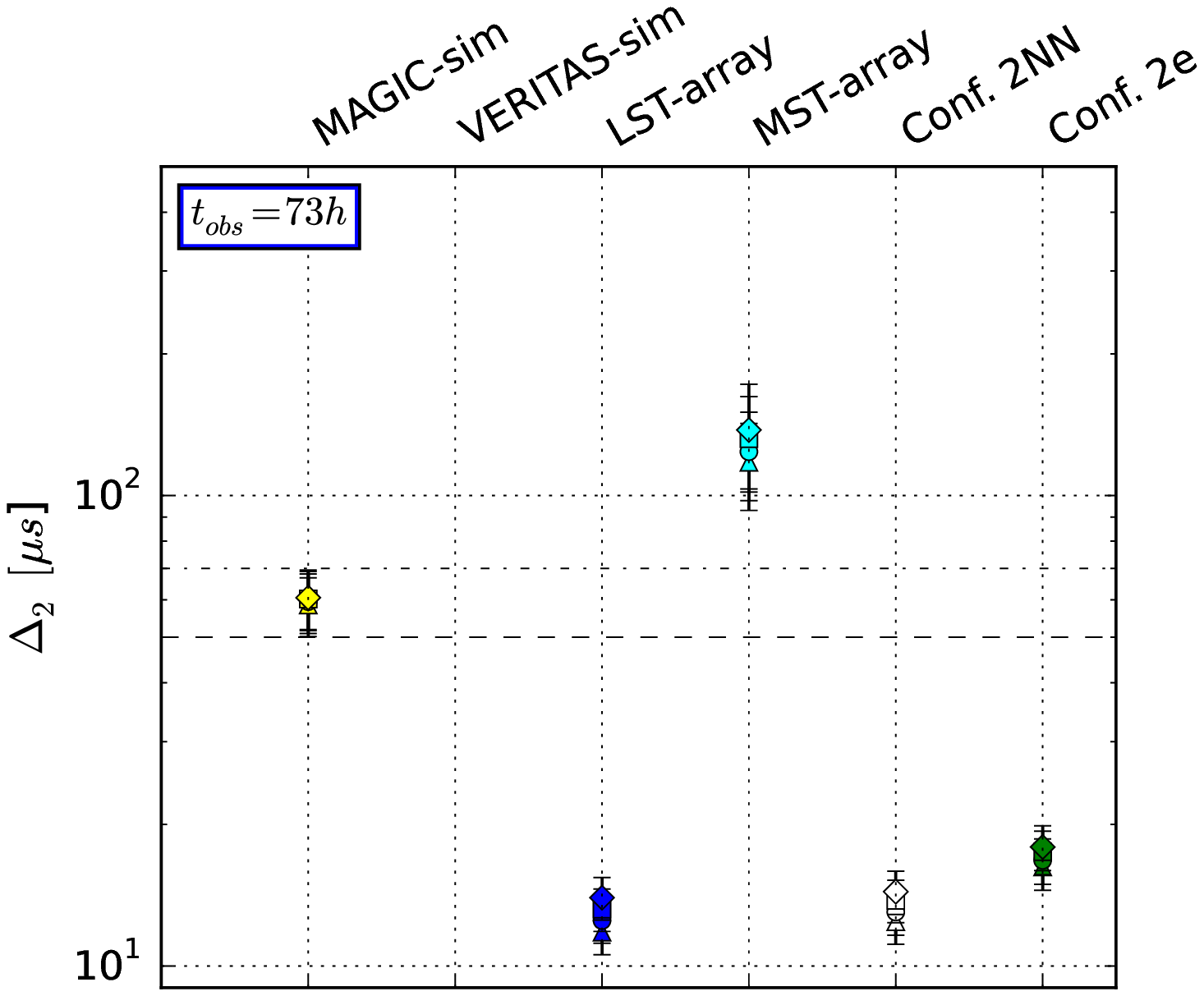}
	\caption{Same as Fig. \ref{fig:D2_ind_Prod2_Array_New}, but restricted to the low energy range (0.04--0.1 TeV). \textit{(A color version of this figure is available on-line).}}
	\label{fig:D2_ind_Prod2_Low_Array}
\end{figure}

\begin{figure}
	\centering
	\includegraphics[width=0.5\textwidth]{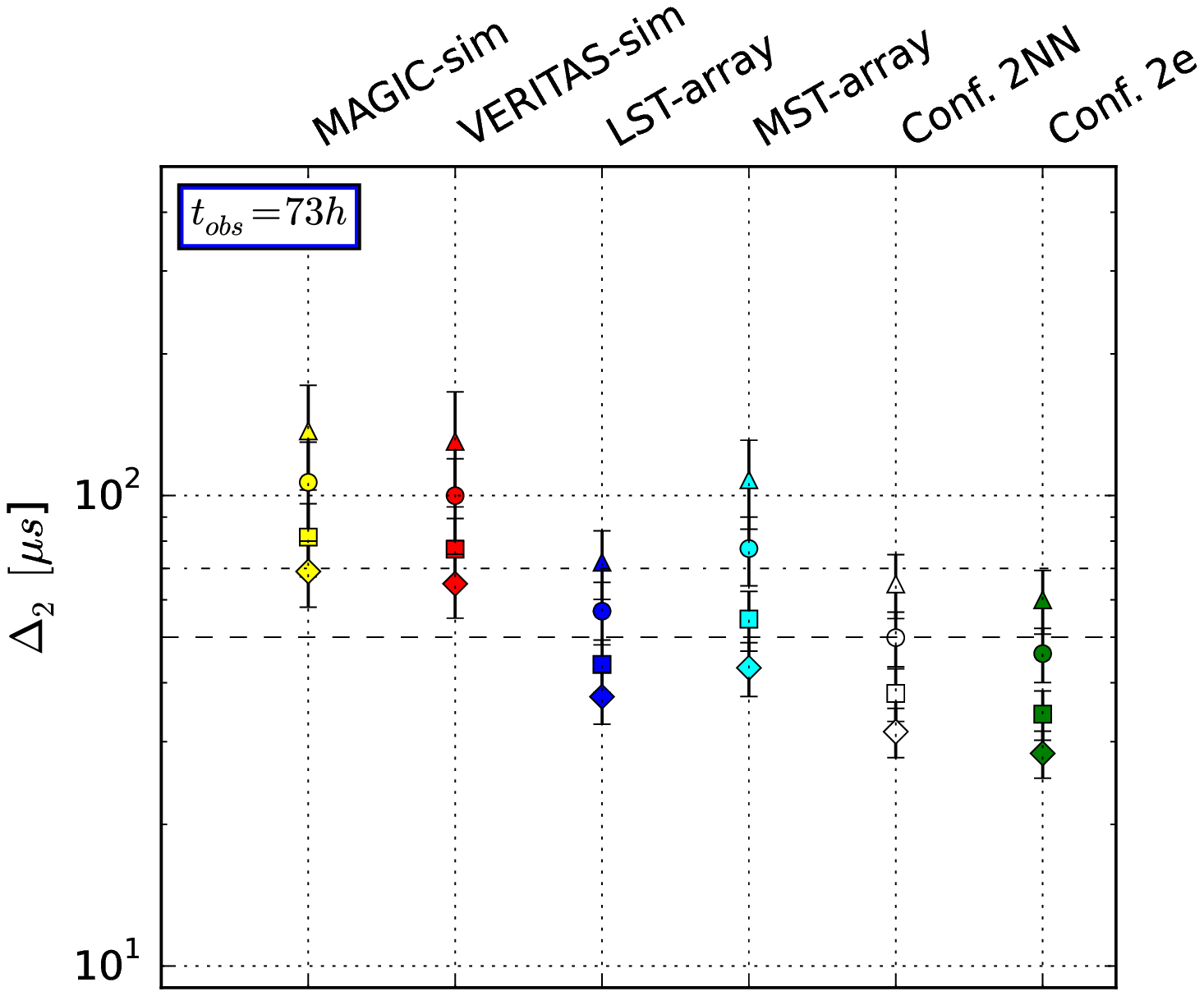} 
	\caption{Same as Fig. \ref{fig:D2_ind_Prod2_Array_New}, but restricted to the mid energy range (0.1--1 TeV). \textit{(A color version of this figure is available on-line).}}
	\label{fig:D2_ind_Prod2_Mid_Array}
\end{figure}

\begin{figure}
	\centering
	\includegraphics[width=0.5\textwidth]{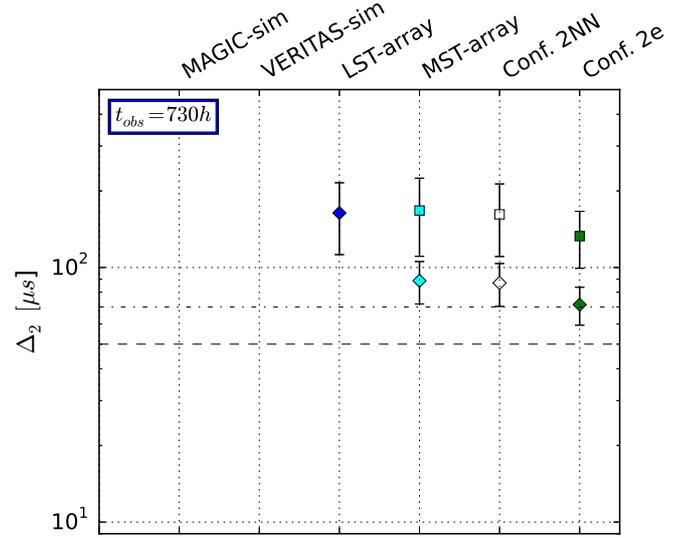} 
	\caption{Same as Fig. \ref{fig:D2_ind_Prod2_Array_New}, but restricted to the high energy range (1--10 TeV) and for the observing times of $t_{\rm obs} = 730$ hours. \textit{(A color version of this figure is available on-line).}}
	\label{fig:D2_ind_Prod2_High_Array}
\end{figure}

We also performed similar simulations using different values of the spectral index $\Gamma$ (3.0, 3.2, 3.5, 3.8) of the Crab pulsar. For each $\Gamma$ we calculate the normalization factor $N_0$, stating that the flux in the energy range from 0.05 to 0.4 TeV is equal to that obtained with MAGIC in the same energy interval \citep{Aleksic2012}. The values of $\Delta_2$ resulting from the simulations of 73-hour observations for all configurations from Table \ref{tab:1} in their full energy range and for simulations restricted to the low (0.04--0.1 TeV) and mid (0.1--1 TeV) energy ranges are shown in Figs. \ref{fig:D2_ind_Prod2_Array_New}, \ref{fig:D2_ind_Prod2_Low_Array} and \ref{fig:D2_ind_Prod2_Mid_Array}, respectively. The results of 730-hour observations at energies from 1 to 10 TeV are reported in Fig. \ref{fig:D2_ind_Prod2_High_Array}.

\subsection{VHE timing analysis}\label{subs:4.1} 
The quality of the pulse shape obtained with the LST-array suggests that a VHE timing analysis of the Crab pulsar, similar to that performed at lower energies (radio/optical/X-ray/low-energy gamma-ray bands), is possible with CTA. We attempted to perform such an analysis of the simulated pulse profile using an approach similar to that discussed in \citet{Germana2012} and \citet{Zampieri2014}.

The time required for the LST-array to achieve statistically significant detection of the pulsar period and the pulse shape is $\sim$1 hour. From the period derivative of the Crab pulsar, one can estimate the phase drift during time $\Delta t$ as $\Delta\phi_{\rm drift}\approx\dot{\nu}\Delta t^2/2$. Assuming $\dot{\nu}\approx -3.7\times10^{-10}$ s$^{-2}$ (see e.g. \citealt{Zampieri2014}) and $\Delta t= 3$ hours, we find $|\Delta\phi_{\rm drift}| \approx 0.02$,
value comparable to the bin size adopted here ($1/N_{\rm bins}$). Therefore, no more than three consecutive 1-hour observations can be performed without a significant phase drift of the Crab pulsar pulse profile.

We investigated the phase drift of the interpulse P2 (more prominent at VHE than the main pulse), measured  with short LSTs observations during a number of consecutive nights. Such measurements require an accurate initial estimate of the Crab pulsar period.  For each night, we simulate three 1-hour observations assuming a parabolic law for the phase drift:
\begin{equation}
	\psi(t) = \phi_0 + a(t-t_0) + b(t-t_0)^2 \,,
	\label{eq:8}
\end{equation}
where $\phi_0$ is the pulsar phase at $t_0$, $a=(\nu_0-\nu_{\rm init})$ is the difference between the rotational frequency of the pulsar $\nu_0$ at $t_0$ and a reference frequency $\nu_{\rm init}$. $b$ is equal to $\dot{\nu_0}/2$, where $\nu_0$ is the rotational frequency first derivative at $t_0$. $\nu_{\rm init}$ is the reference frequency used to fold the data. In our analysis we assume that the difference between $\nu_0$ and $\nu_{\rm init}$ is of the order of $10^{-5}$ s$^{-1}$ and that $\dot{\nu_0}\simeq -3.7\times10^{-10}$ s$^{-2}$ (see e.g. \citet{Germana2012} or \citealt{Zampieri2014}). The adopted values of $\phi_0$, $a$ and $b$ are reported in the first line of Table \ref{tab:2}. In order to accurately fold real data, it will require changing reference frequency each night. It is possible to reduce the phase measurements to a single reference frequency using the method described in \citet{Zampieri2014}.

\begin{table*}
\caption{Parameters of the parabolic fit of the phase drift. The first line shows the assumed spin-down law, while the second and third lines contain the best fitting values of the parameters obtained from a fit of the simulated spin-down for two different observing intervals (2 nights and 3 nights; see text for details).}
\label{tab:2}
\centering
\begin{tabular}{l c c c}
\hline\hline
 & $\phi_0$ & $a$ & $b$ \\
 & & ($10^{-5} ~\mathrm{s}^{-1}$) 	& ($10^{-10} ~\mathrm{s}^{-2}$) \\
\hline
$\psi$\, 			& 0.394 				&  1.0			& -1.85  				\\ 
\hline
$\psi_{\rm 2d}$	& 0.386$\pm 0.004$	& $1.05 \pm 0.06$ 	& -1.90$\pm 0.06$		\\
$\psi_{\rm 3d}$	& 0.390$\pm 0.002$	& $0.995\pm 0.006$& -1.847$\pm 0.003$	\\
\hline
\end{tabular}
\end{table*}

\begin{figure}
	\centering
	\includegraphics[width=0.50\textwidth]{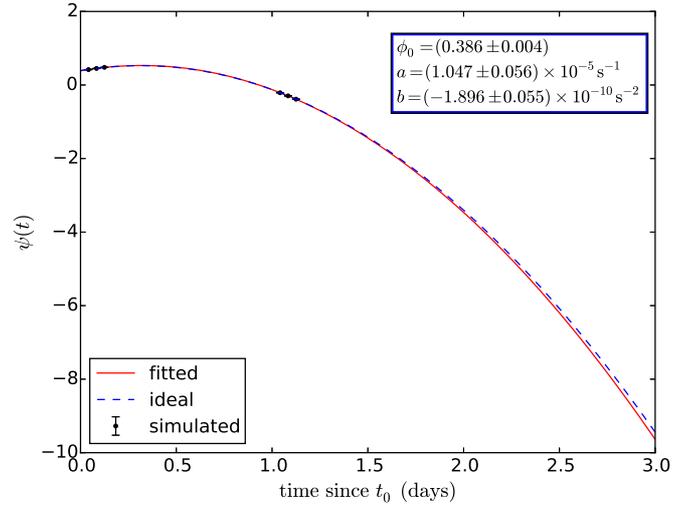}
	\caption{Simulated phase drift of the VHE interpulse of the Crab pulsar observed with the LST-array over 2 nights. The blue dashed line is the assumed spin-down law. The red solid line is the best-fit parabola. Black dots are the simulated data. \textit{(A color version of this figure is available on-line)}.}
	\label{fig:Psi1}
\end{figure}

\begin{figure}
	\centering
	\includegraphics[width=0.50\textwidth]{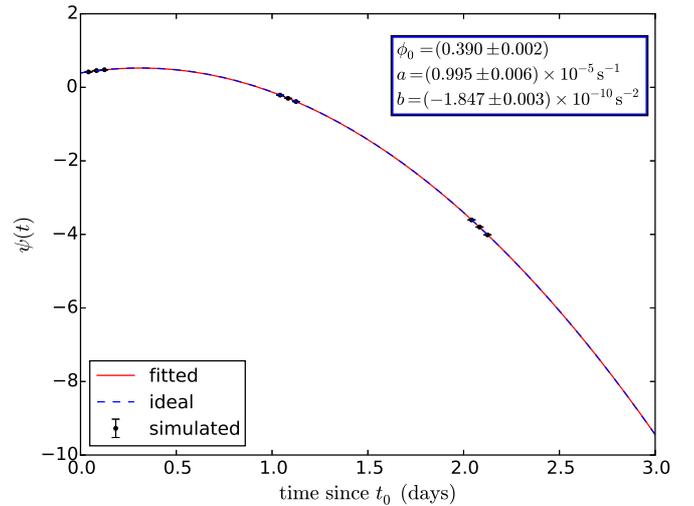}
	\caption{Same as Fig. \ref{fig:Psi1} for observations covering 3 nights. \textit{(A color version of this figure is available on-line)}.}
	\label{fig:Psi2}
\end{figure}

Following the procedure described in Sect. \ref{sec:2}, for each observation we simulated the pulse profile detected with the LST-array\footnote{For these simulations we assume a power-law spectrum for the Crab pulsar with $N_0 = 13.0 \times10^{-11}$ TeV$^{-1}$ cm$^{-2}$ s$^{-1}$ and $\Gamma = 3.57$.} and obtained the phase drift $\psi_i$ and error $\sigma_i$ of the interpulse in the $i$-th observation. Typically, $\sigma_i \sim$120 $\mu$s. We then simulated 2 and 3 nights of observations of the Crab pulsar (see Figs. \ref{fig:Psi1} and \ref{fig:Psi2}). The positions of the interpulse, derived from the simulated observations and reduced to the same reference frequency, were then fitted with the parabolic law in Eq. (\ref{eq:8}). Best fit coefficients and corresponding errors are reported in Table \ref{tab:2} (second and third lines). From these values it is possible to estimate the accuracy in determining the time of arrival of the interpulse, which is $\sim$140 $\mu$s after 2 nights and $\sim$80 $\mu$s after 3 nights (assuming $P(t_0)=0.03362$ s, which is the approximate rotational period of the Crab pulsar at $t_0$, see e.g. \citealt{Zampieri2014}). Increasing the number of observing nights does not improve significantly the accuracy of the fitting parameters (e.g. the accuracy in determining the position of the interpulse is $\sim$65 $\mu$s after 7 nights).

For the MSTs and SSTs this type of measurements of the phase drift of the interpulse is not feasible because detecting a pulse profile requires more than one observing night.

\section{Discussion}\label{sec:5}
We performed simulations of the VHE gamma-ray pulse profile of the Crab pulsar for different configurations of CTA and the ASTRI mini-array, therefore in intrinsically different energy ranges and with different  observing times. The LSTs, MSTs and SSTs will probe different spectral regions (from a few tens of GeV up to several tens of TeV) with different sensitivities \citep{Bernlohr2013}. For all simulated pulse profiles we determined the uncertainty $\Delta_2$ in the position of the interpulse.

As shown in Fig. \ref{fig:Final_P2_A}, the LST-array reaches the same accuracy as MAGIC ($\Delta_2 \sim 50$ $\mu$s) but in a much shorter observation time ($0.1\times t_{\rm obs}^{\rm M}$). 73 hours of observations with the same configuration lead to an uncertainty in the peak position of 14 $\mu$s. Observations with the northern configuration Conf. 2NN give an improvement by a factor $\sim$3 in accuracy as compared to MAGIC (with 73 hours of observing time; see again Fig. \ref{fig:Final_P2_A}). Similar values of $\Delta_2$ are obtained for the full-energy-range configurations Conf. 2e, which contains all three types of telescopes. These results are slightly worse than that attainable with the LST-array because of the lower background contamination and larger effective area of the latter configuration in the energy range below a few hundred GeV, where the Crab pulsar is easier to detect due to its steep spectrum.

The quality of the measured pulse shape decreases significantly above 1 TeV again because of the steeply falling spectrum. In spite of the increase in the effective area with energy, the value of $\Delta_2$ for the MST-array measured above 100 GeV is nearly the same as that of MAGIC above 50 GeV. This is even more the case for the high-energy-range array of SSTs. We estimated, that only with a $10^4$ times longer observing time ($\sim7\times 10^5$ hours), will an array of 72 SSTs return a value of $\Delta_2$ (above a few TeV) comparable with that of VERITAS above 100 GeV. Similar conclusions can be drawn for the ASTRI mini-array, which contains only nine SSTs. Pure detection of the pulsed emission in this case would require $\sim$$10^6$ hours and is thus not achievable. 

We note that all estimates depend on the values of the effective area and energy threshold. Modifications of the telescopes design and arrays configurations can affect them and, therefore, change the results presented here.

Clearly, these results are very sensitive also to the VHE spectral index $\Gamma$ of the Crab pulsar. We performed similar simulations assuming different values of $\Gamma$ (3.0, 3.2, 3.5, 3.8) in the full (0.04--160 TeV), low (0.04--0.1 TeV), mid (0.1--1 TeV) and high (1--10 TeV) energy ranges (see Figs. \ref{fig:D2_ind_Prod2_Array_New}--\ref{fig:D2_ind_Prod2_High_Array}). Below 100 GeV the best $\Delta_2$ is provided by the LST-array. In the energy range 0.1--1 TeV the most accurate values of $\Delta_2$ are obtained with Conf. 2NN -- the northern CTA installation, which contains LSTs and MSTs -- and with Conf. 2e, which comprises all three types of telescopes (LSTs, MSTs, SSTs). Even if the spectrum of the Crab pulsar is rather steep ($\Gamma=3.8$) in this energy range, CTA will be able to reach an accuracy $\Delta_2\sim60$ $\mu$s in 73 hours (see green triangle in Fig. \ref{fig:D2_ind_Prod2_Mid_Array}). However, above 1 TeV only with 10 times longer observations (730 hours) and assuming a hard spectrum for the Crab pulsar ($\Gamma=3.0$), will Conf. 2e perform an accurate measurement of the position of the interpulse P2 ($\Delta_2=70~\mu$s).

Theoretical models predict different spectral behaviors of isolated pulsars at VHE. \citet{Aharonian2012} presented a mechanism of VHE gamma ray production through inverse Compton (IC) scattering of X-ray photons on relativistic electrons, accelerated in a region located far beyond the light cylinder of the neutron star (from $20R_L$ to $50R_L$, where $R_L$ is a light-cylinder radius). This model predicts a cut-off at 500 GeV. \citet{Lyutikov2012} showed that ultraviolet and X-ray photons produced in the inner magnetosphere can be up-scattered to VHE in the outer magnetosphere and produce a spectral tail extending up to $\sim$15 TeV (if the accelerating electric field is 100 times lower than the magnetic field of the neutron star, and the curvature radius of the order of the light cylinder radius $R_L$). IC scattering on a relativistic electron-positron pair plasma accelerated in annular or core gap regions predicts a VHE gamma-ray spectrum reaching 400 GeV \citep{Du2012}.

In order to determine the VHE folded profiles needed for the present analysis, an accurate knowledge of the Crab pulsar ephemerids is required. This can be obtained from simultaneous observations at lower energies (e.g. radio, optical). However, we also investigated the possibility of performing an independent phase timing analysis at VHE only with CTA, using observations spread over several nights. The strategy is similar to that adopted in \citet{Germana2012} and \citet{Zampieri2014}. We simulated 2- and 3-night observations with the LST-array (three 1-hour exposures each night). The accuracy on the time of arrival of the interpulse is $\sim$140 $\mu$s and $\sim$80 $\mu$s for observations covering 2 or 3 nights, respectively. Resulting values are worse than those derived from a fit of the pulse profile obtained folding together all observations (using known ephemerides). Thus, although an independent VHE timing analysis based on short repeated observations appears to be feasible with the LSTs, for the sake of measuring $\Delta_2$ the obtained results are less accurate.

\section{Conclusions}\label{sec:6}
The energy spectrum and pulse profile at VHE are crucial ingredients for any comprehensive theory of pulsar emission. Different mechanisms for particle acceleration and VHE gamma-ray emission have been proposed \citep[see][]{Aharonian2012, Lyutikov2012,Mochol2015arx}. Some models \citep[see][]{Bai2010_2} can predict the shape of the pulse profile and yield different time shifts between the position of the peaks at VHE and in the radio band. Because of its better sensitivity and wider energy range CTA will provide crucial input for the theory. Together with the full CTA, the LSTs- and MSTs-arrays will provide an accurate measurement of the time of arrival of the peaks at VHE, and will then allow us to determine its shift with respect to simultaneous measurements in other energy bands (radio, optical, X-rays, low-energy gamma rays \citep{Abdo2010}). In this respect, it would be important that presently on-going monitoring programs of the Crab pulsar at different wavelengths (e.g. that of Jodrell Bank in the radio) continue to operate.

Configurations containing LSTs/MSTs (with threshold energy $E_{\rm thr}$ equal to 0.04/0.16 TeV) will be able to measure more detailed features in the VHE pulse profile, which will further constrain the emission region and emission mechanism of pulsars. Any potential phase shift between the LST- and MST-arrays significantly larger than $\sim$270 $\mu$s will also be detectable. On the other hand, extrapolating the power-law spectral shape inferred at lower-energies, an accurate determination of the pulse profile of the Crab pulsar with the high-energy SSTs ($E_{\rm thr} = 1$ TeV) is essentially not possible.

\section*{Acknowledgements}
This work was partially supported by the ASTRI ``Flagship Project'' financed by the Italian Ministry of Education, University, and Research (MIUR) and led by the Italian National Institute of Astrophysics (INAF). We acknowledge partial support by the MIUR Bando PRIN 2009 and TeChe.it 2014 Special Grants. We also acknowledge support from the Brazilian Funding Agency FAPESP (Grant 2013/10559-5) and from the South African Department of Science and Technology through Funding Agreement 0227/2014 for the South African Gamma-Ray Astronomy Programme. We gratefully acknowledge support from the agencies and organizations listed under Funding Agencies at \url{http://www.cta-observatory.org/} and from the University of Padova.

\bibliographystyle{mnras} 

\begin{thebibliography}{}
\makeatletter
\relax
\def\mn@urlcharsother{\let\do\@makeother \do\$\do\&\do\#\do\^\do\_\do\%\do\~}
\def\mn@doi{\begingroup\mn@urlcharsother \@ifnextchar [ {\mn@doi@}
  {\mn@doi@[]}}
\def\mn@doi@[#1]#2{\def\@tempa{#1}\ifx\@tempa\@empty \href
  {http://dx.doi.org/#2} {doi:#2}\else \href {http://dx.doi.org/#2} {#1}\fi
  \endgroup}
\def\mn@eprint#1#2{\mn@eprint@#1:#2::\@nil}
\def\mn@eprint@arXiv#1{\href {http://arxiv.org/abs/#1} {{\tt arXiv:#1}}}
\def\mn@eprint@dblp#1{\href {http://dblp.uni-trier.de/rec/bibtex/#1.xml}
  {dblp:#1}}
\def\mn@eprint@#1:#2:#3:#4\@nil{\def\@tempa {#1}\def\@tempb {#2}\def\@tempc
  {#3}\ifx \@tempc \@empty \let \@tempc \@tempb \let \@tempb \@tempa \fi \ifx
  \@tempb \@empty \def\@tempb {arXiv}\fi \@ifundefined
  {mn@eprint@\@tempb}{\@tempb:\@tempc}{\expandafter \expandafter \csname
  mn@eprint@\@tempb\endcsname \expandafter{\@tempc}}}

\bibitem[\protect\citeauthoryear{{Abdo} et~al.,}{{Abdo}
  et~al.}{2010}]{Abdo2010}
{Abdo} A.~A.,  et~al., 2010, \mn@doi [\apj] {10.1088/0004-637X/708/2/1254},
  \href {http://adsabs.harvard.edu/abs/2010ApJ...708.1254A} {708, 1254}

\bibitem[\protect\citeauthoryear{{Abramowski} et~al.,}{{Abramowski}
  et~al.}{2014}]{Abramowski2014}
{Abramowski} A.,  et~al., 2014, \mn@doi [\aap] {10.1051/0004-6361/201323013},
  \href {http://adsabs.harvard.edu/abs/2014A%26A...562L...4H} {562, L4}

\bibitem[\protect\citeauthoryear{{Acharya} et~al.,}{{Acharya}
  et~al.}{2013}]{Acharya2013}
{Acharya} B.~S.,  et~al., 2013, \mn@doi [Astroparticle Physics]
  {10.1016/j.astropartphys.2013.01.007}, \href
  {http://adsabs.harvard.edu/abs/2013APh....43....3C} {43, 3}

\bibitem[\protect\citeauthoryear{{Actis} et~al.,}{{Actis}
  et~al.}{2011}]{CTA2011}
{Actis} M.,  et~al., 2011, \mn@doi [Experimental Astronomy]
  {10.1007/s10686-011-9247-0}, \href
  {http://adsabs.harvard.edu/abs/2011ExA....32..193A} {32, 193}

\bibitem[\protect\citeauthoryear{{Aharonian} et~al.,}{{Aharonian}
  et~al.}{2004}]{Aharonian2004}
{Aharonian} F.,  et~al., 2004, \mn@doi [\apj] {10.1086/423931}, \href
  {http://adsabs.harvard.edu/abs/2004ApJ...614..897A} {614, 897}

\bibitem[\protect\citeauthoryear{{Aharonian} et~al.,}{{Aharonian}
  et~al.}{2006}]{Aharonian2006_2}
{Aharonian} F.,  et~al., 2006, \mn@doi [\aap] {10.1051/0004-6361:20065351},
  \href {http://adsabs.harvard.edu/abs/2006A%26A...457..899A} {457, 899}

\bibitem[\protect\citeauthoryear{{Aharonian}, {Bogovalov}  \&
  {Khangulyan}}{{Aharonian} et~al.}{2012}]{Aharonian2012}
{Aharonian} F.~A.,  {Bogovalov} S.~V.,   {Khangulyan} D.,  2012, \mn@doi [\nat]
  {10.1038/nature10793}, \href
  {http://adsabs.harvard.edu/abs/2012Natur.482..507A} {482, 507}

\bibitem[\protect\citeauthoryear{{Aleksi{\'c}} et~al.,}{{Aleksi{\'c}}
  et~al.}{2011}]{Aleksic2011}
{Aleksi{\'c}} J.,  et~al., 2011, \mn@doi [\apj] {10.1088/0004-637X/742/1/43},
  \href {http://adsabs.harvard.edu/abs/2011ApJ...742...43A} {742, 43}

\bibitem[\protect\citeauthoryear{{Aleksi{\'c}} et~al.,}{{Aleksi{\'c}}
  et~al.}{2012a}]{Aleksic2012_2}
{Aleksi{\'c}} J.,  et~al., 2012a, \mn@doi [Astroparticle Physics]
  {10.1016/j.astropartphys.2011.11.007}, \href
  {http://adsabs.harvard.edu/abs/2012APh....35..435A} {35, 435}

\bibitem[\protect\citeauthoryear{{Aleksi{\'c}} et~al.,}{{Aleksi{\'c}}
  et~al.}{2012b}]{Aleksic2012}
{Aleksi{\'c}} J.,  et~al., 2012b, \mn@doi [\aap] {10.1051/0004-6361/201118166},
  \href {http://adsabs.harvard.edu/abs/2012A%26A...540A..69A} {540, A69}

\bibitem[\protect\citeauthoryear{{Aleksi{\'c}} et~al.,}{{Aleksi{\'c}}
  et~al.}{2014}]{Aleksic2014_2}
{Aleksi{\'c}} J.,  et~al., 2014, \mn@doi [\aap] {10.1051/0004-6361/201423664},
  \href {http://adsabs.harvard.edu/abs/2014A%26A...565L..12A} {565, L12}

\bibitem[\protect\citeauthoryear{{Aleksi{\'c}} et~al.,}{{Aleksi{\'c}}
  et~al.}{2015}]{Aleksic2015}
{Aleksi{\'c}} J.,  et~al., 2015, \mn@doi [Journal of High Energy Astrophysics]
  {10.1016/j.jheap.2015.01.002}, \href
  {http://adsabs.harvard.edu/abs/2015JHEAp...5...30A} {5, 30}

\bibitem[\protect\citeauthoryear{{Aliu} et~al.,}{{Aliu}
  et~al.}{2008}]{Aliu2008}
{Aliu} E.,  et~al., 2008, \mn@doi [Science] {10.1126/science.1164718}, \href
  {http://adsabs.harvard.edu/abs/2008Sci...322.1221A} {322, 1221}

\bibitem[\protect\citeauthoryear{{Aliu} et~al.,}{{Aliu}
  et~al.}{2011}]{Aliu2011}
{Aliu} E.,  et~al., 2011, \mn@doi [Science] {10.1126/science.1208192}, \href
  {http://adsabs.harvard.edu/abs/2011Sci...334...69V} {334, 69}

\bibitem[\protect\citeauthoryear{{Bai} \& {Spitkovsky}}{{Bai} \&
  {Spitkovsky}}{2010}]{Bai2010_2}
{Bai} X.-N.,  {Spitkovsky} A.,  2010, \mn@doi [\apj]
  {10.1088/0004-637X/715/2/1282}, \href
  {http://adsabs.harvard.edu/abs/2010ApJ...715.1282B} {715, 1282}

\bibitem[\protect\citeauthoryear{{Bernl{\"o}hr} et~al.,}{{Bernl{\"o}hr}
  et~al.}{2013}]{Bernlohr2013}
{Bernl{\"o}hr} K.,  et~al., 2013, \mn@doi [Astroparticle Physics]
  {10.1016/j.astropartphys.2012.10.002}, \href
  {http://adsabs.harvard.edu/abs/2013APh....43..171F} {43, 171}

\bibitem[\protect\citeauthoryear{{Cusumano} et~al.,}{{Cusumano}
  et~al.}{2012}]{Cusumano2012}
{Cusumano} G.,  et~al., 2012, \mn@doi [\aap] {10.1051/0004-6361/201219968},
  \href {http://adsabs.harvard.edu/abs/2012A%26A...548A..28C} {548, A28}

\bibitem[\protect\citeauthoryear{{Du}, {Qiao}  \& {Wang}}{{Du}
  et~al.}{2012}]{Du2012}
{Du} Y.~J.,  {Qiao} G.~J.,   {Wang} W.,  2012, \mn@doi [\apj]
  {10.1088/0004-637X/748/2/84}, \href
  {http://adsabs.harvard.edu/abs/2012ApJ...748...84D} {748, 84}

\bibitem[\protect\citeauthoryear{{German{\`a}} et~al.,}{{German{\`a}}
  et~al.}{2012}]{Germana2012}
{German{\`a}} C.,  et~al., 2012, \mn@doi [\aap] {10.1051/0004-6361/201118754},
  \href {http://adsabs.harvard.edu/abs/2012A%26A...548A..47G} {548, A47}

\bibitem[\protect\citeauthoryear{{Golden}, {Shearer}, {Redfern}, {Beskin},
  {Neizvestny}, {Neustroev}, {Plokhotnichenko}  \& {Cullum}}{{Golden}
  et~al.}{2000}]{Golden2000}
{Golden} A.,  {Shearer} A.,  {Redfern} R.~M.,  {Beskin} G.~M.,  {Neizvestny}
  S.~I.,  {Neustroev} V.~V.,  {Plokhotnichenko} V.~L.,   {Cullum} M.,  2000,
  \aap, \href {http://adsabs.harvard.edu/abs/2000A%26A...363..617G} {363, 617}

\bibitem[\protect\citeauthoryear{{Grube}}{{Grube}}{2008}]{Grube2008}
{Grube} J.,  2008, International Cosmic Ray Conference, \href
  {http://adsabs.harvard.edu/abs/2008ICRC....2..691G} {2, 691}

\bibitem[\protect\citeauthoryear{{Hinton}, {Hermann}, {Kr{\"o}tz}  \&
  {Funk}}{{Hinton} et~al.}{2006}]{Hinton2006}
{Hinton} J.,  {Hermann} G.,  {Kr{\"o}tz} P.,   {Funk} S.,  2006, \mn@doi
  [Astroparticle Physics] {10.1016/j.astropartphys.2006.04.008}, \href
  {http://adsabs.harvard.edu/abs/2006APh....26...22H} {26, 22}

\bibitem[\protect\citeauthoryear{{Hobbs}, {Lyne}, {Kramer}, {Martin}  \&
  {Jordan}}{{Hobbs} et~al.}{2004}]{Hobbs2004}
{Hobbs} G.,  {Lyne} A.~G.,  {Kramer} M.,  {Martin} C.~E.,   {Jordan} C.,  2004,
  \mn@doi [\mnras] {10.1111/j.1365-2966.2004.08157.x}, \href
  {http://adsabs.harvard.edu/abs/2004MNRAS.353.1311H} {353, 1311}

\bibitem[\protect\citeauthoryear{{Kieda D.~B.~ for the VERITAS
  Collaboration}}{{Kieda D.~B.~ for the VERITAS
  Collaboration}}{2013}]{Kieda2013arx}
{Kieda D.~B.~ for the VERITAS Collaboration} 2013, [arXiv:1308.4849], \href
  {http://adsabs.harvard.edu/abs/2013arXiv1308.4849D} {}

\bibitem[\protect\citeauthoryear{{Kirsch} et~al.,}{{Kirsch}
  et~al.}{2006}]{Kirsch2006}
{Kirsch} M.~G.~F.,  et~al., 2006, \mn@doi [\aap] {10.1051/0004-6361:20054783},
  \href {http://adsabs.harvard.edu/abs/2006A%26A...453..173K} {453, 173}

\bibitem[\protect\citeauthoryear{{Kuiper}, {Hermsen}, {Cusumano}, {Diehl},
  {Sch{\"o}nfelder}, {Strong}, {Bennett}  \& {McConnell}}{{Kuiper}
  et~al.}{2001}]{Kuiper2001}
{Kuiper} L.,  {Hermsen} W.,  {Cusumano} G.,  {Diehl} R.,  {Sch{\"o}nfelder} V.,
   {Strong} A.,  {Bennett} K.,   {McConnell} M.~L.,  2001, \mn@doi [\aap]
  {10.1051/0004-6361:20011256}, \href
  {http://adsabs.harvard.edu/abs/2001A%26A...378..918K} {378, 918}

\bibitem[\protect\citeauthoryear{{La Palombara} et~al.,}{{La Palombara}
  et~al.}{2014}]{LaPalombara2014}
{La Palombara} N.,  et~al., 2014, in {Giani} S.,  {et al.} eds, Astroparticle,
  Particle, Space Physics and Detectors for Physics Applications - Proceedings
  of the 14th ICATPP Conference. pp 754--758 (\mn@eprint {arXiv} {1405.4187}),
  \mn@doi{10.1142/9789814603164_0119}

\bibitem[\protect\citeauthoryear{{Lyutikov}, {Otte}  \& {McCann}}{{Lyutikov}
  et~al.}{2012}]{Lyutikov2012}
{Lyutikov} M.,  {Otte} N.,   {McCann} A.,  2012, \mn@doi [\apj]
  {10.1088/0004-637X/754/1/33}, \href
  {http://adsabs.harvard.edu/abs/2012ApJ...754...33L} {754, 33}

\bibitem[\protect\citeauthoryear{{MAGIC Collaboration} et~al.,}{{MAGIC
  Collaboration} et~al.}{2015}]{Ahnen2015arx}
{MAGIC Collaboration} et~al., 2015, preprint, \href
  {http://adsabs.harvard.edu/abs/2015arXiv151007048M} {} (\mn@eprint {arXiv}
  {1510.07048})

\bibitem[\protect\citeauthoryear{{Manchester}, {Hobbs}, {Teoh}  \&
  {Hobbs}}{{Manchester} et~al.}{2005}]{Manchester2005}
{Manchester} R.~N.,  {Hobbs} G.~B.,  {Teoh} A.,   {Hobbs} M.,  2005, \mn@doi
  [\aj] {10.1086/428488}, \href
  {http://adsabs.harvard.edu/abs/2005AJ....129.1993M} {129, 1993}

\bibitem[\protect\citeauthoryear{{Masterson} \& {CAT
  Collaboration}}{{Masterson} \& {CAT Collaboration}}{2001}]{Masterson2001}
{Masterson} C.,  {CAT Collaboration} 2001, in {Aharonian} F.~A.,  {V{\"o}lk}
  H.~J.,  eds,  American Institute of Physics Conference Series Vol. 558,
  American Institute of Physics Conference Series. p.~753

\bibitem[\protect\citeauthoryear{{Mineo}, {Ferrigno}, {Foschini}, {Segreto},
  {Cusumano}, {Malaguti}, {Di Cocco}  \& {Labanti}}{{Mineo}
  et~al.}{2006}]{Mineo2006}
{Mineo} T.,  {Ferrigno} C.,  {Foschini} L.,  {Segreto} A.,  {Cusumano} G.,
  {Malaguti} G.,  {Di Cocco} G.,   {Labanti} C.,  2006, \mn@doi [\aap]
  {10.1051/0004-6361:20054305}, \href
  {http://adsabs.harvard.edu/abs/2006A%26A...450..617M} {450, 617}

\bibitem[\protect\citeauthoryear{{Mochol} \& {Petri}}{{Mochol} \&
  {Petri}}{2015}]{Mochol2015arx}
{Mochol} I.,  {Petri} J.,  2015, [arXiv:1501.07123], \href
  {http://adsabs.harvard.edu/abs/2015arXiv150107123M} {}

\bibitem[\protect\citeauthoryear{{Oosterbroek} et~al.,}{{Oosterbroek}
  et~al.}{2008}]{Oosterbroek2008}
{Oosterbroek} T.,  et~al., 2008, \mn@doi [\aap] {10.1051/0004-6361:200809751},
  \href {http://adsabs.harvard.edu/abs/2008A%26A...488..271O} {488, 271}

\bibitem[\protect\citeauthoryear{{Pellizzoni} et~al.,}{{Pellizzoni}
  et~al.}{2009}]{Pellizzoni2009}
{Pellizzoni} A.,  et~al., 2009, \mn@doi [\apj] {10.1088/0004-637X/691/2/1618},
  \href {http://adsabs.harvard.edu/abs/2009ApJ...691.1618P} {691, 1618}

\bibitem[\protect\citeauthoryear{{Rots}, {Jahoda}  \& {Lyne}}{{Rots}
  et~al.}{2004}]{Rots2004}
{Rots} A.~H.,  {Jahoda} K.,   {Lyne} A.~G.,  2004, \mn@doi [\apjl]
  {10.1086/420842}, \href {http://adsabs.harvard.edu/abs/2004ApJ...605L.129R}
  {605, L129}

\bibitem[\protect\citeauthoryear{{Shearer}, {Stappers}, {O'Connor}, {Golden},
  {Strom}, {Redfern}  \& {Ryan}}{{Shearer} et~al.}{2003}]{Shearer2003}
{Shearer} A.,  {Stappers} B.,  {O'Connor} P.,  {Golden} A.,  {Strom} R.,
  {Redfern} M.,   {Ryan} O.,  2003, \mn@doi [Science]
  {10.1126/science.1084919}, \href
  {http://adsabs.harvard.edu/abs/2003Sci...301..493S} {301, 493}

\bibitem[\protect\citeauthoryear{{Terada} et~al.,}{{Terada}
  et~al.}{2008}]{Terada2008}
{Terada} Y.,  et~al., 2008, \pasj, \href
  {http://adsabs.harvard.edu/abs/2008PASJ...60S..25T} {60, 25}

\bibitem[\protect\citeauthoryear{{Theureau} et~al.,}{{Theureau}
  et~al.}{2005}]{Theureau2005}
{Theureau} G.,  et~al., 2005, \mn@doi [\aap] {10.1051/0004-6361:20047152},
  \href {http://adsabs.harvard.edu/abs/2005A%26A...430..373T} {430, 373}

\bibitem[\protect\citeauthoryear{{Vercellone}, {for The ASTRI Collaboration}
  \& {CTA Consortium}}{{Vercellone} et~al.}{2015}]{Vercellone2015arx}
{Vercellone} S.,  {for The ASTRI Collaboration}  {CTA Consortium} f.~T.,  2015,
  [arXiv:1508.00799], \href {http://adsabs.harvard.edu/abs/2015arXiv150800799V}
  {}

\bibitem[\protect\citeauthoryear{{Weekes} et~al.,}{{Weekes}
  et~al.}{1989}]{Weekes1989}
{Weekes} T.~C.,  et~al., 1989, \mn@doi [\apj] {10.1086/167599}, \href
  {http://adsabs.harvard.edu/abs/1989ApJ...342..379W} {342, 379}

\bibitem[\protect\citeauthoryear{{Zampieri} et~al.,}{{Zampieri}
  et~al.}{2014}]{Zampieri2014}
{Zampieri} L.,  et~al., 2014, \mn@doi [\mnras] {10.1093/mnras/stu136}, \href
  {http://adsabs.harvard.edu/abs/2014MNRAS.439.2813Z} {439, 2813}

\bibitem[\protect\citeauthoryear{{de O{\~n}a-Wilhelmi} et~al.,}{{de
  O{\~n}a-Wilhelmi} et~al.}{2013}]{Ona2013}
{de O{\~n}a-Wilhelmi} E.,  et~al., 2013, \mn@doi [Astroparticle Physics]
  {10.1016/j.astropartphys.2012.08.009}, \href
  {http://adsabs.harvard.edu/abs/2013APh....43..287D} {43, 287}

\makeatother
\end{thebibliography}

\appendix

\appendix
\section{Results for other MC-Prod2 configurations}\label{app:1}
\begin{figure}
	\centering
	\includegraphics[width=0.48\textwidth]{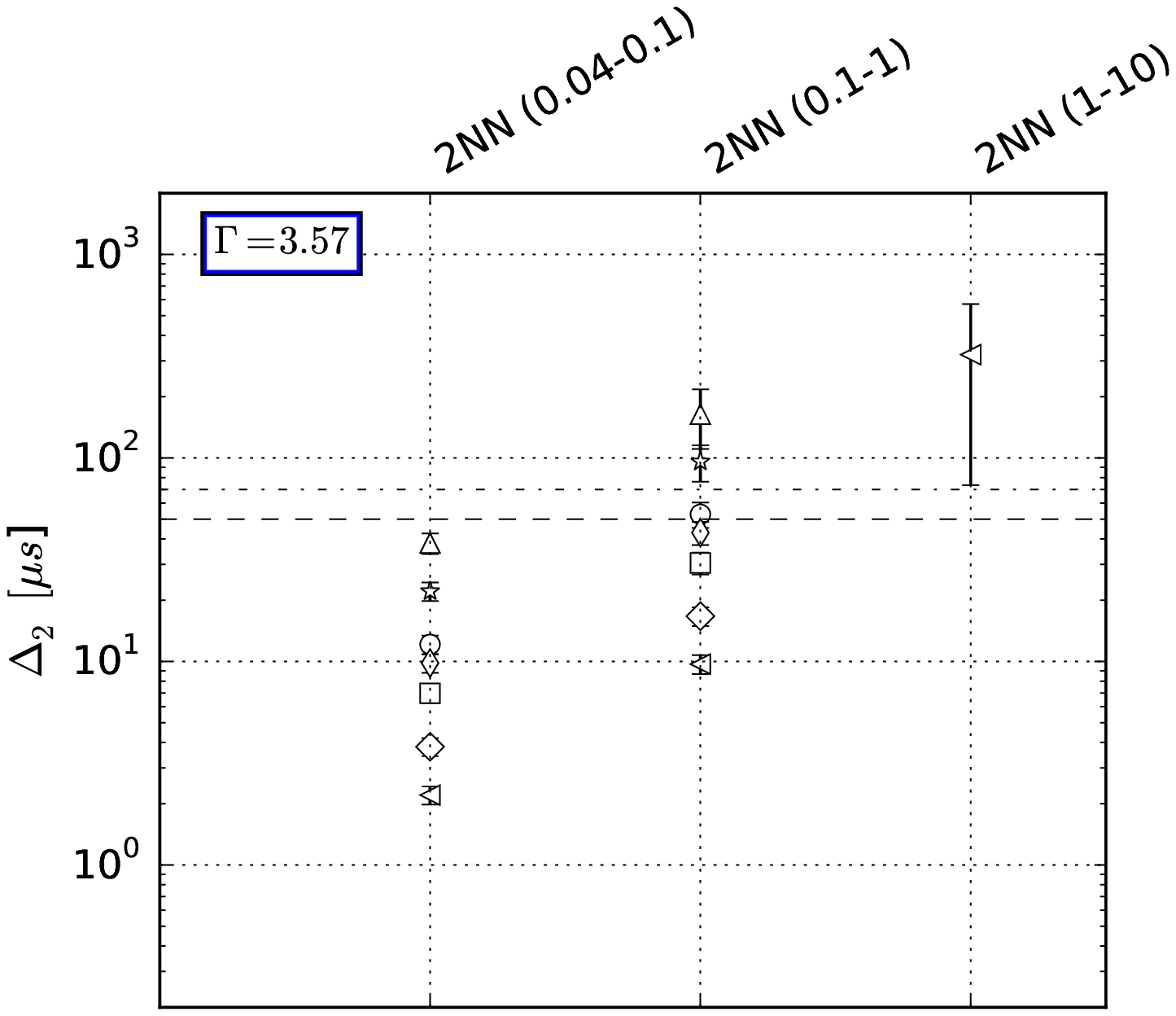}
	\caption{Same as Fig. \ref{fig:Final_P2_A} for different energy ranges of Conf. 2NN. \textit{(A color version of this figure is available on-line)}.}
	\label{fig:Final_P2_ArrEn}
\end{figure}

\begin{figure}
	\centering
	\includegraphics[width=0.48\textwidth]{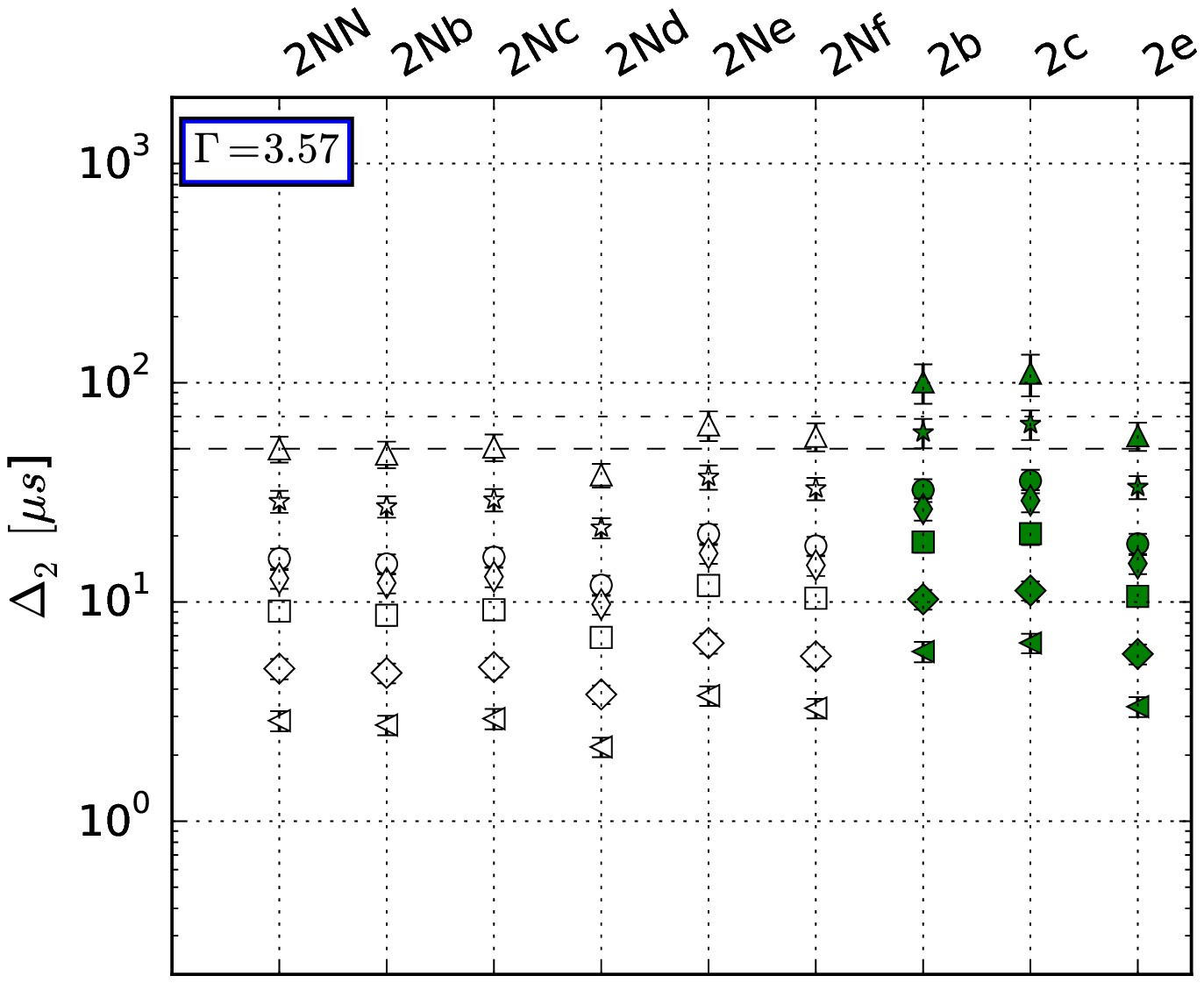} 
	\caption{Same as Fig. \ref{fig:Final_P2_A} for other possible layouts of the CTA-North (Confs. 2NN, 2Nc, 2Ne, 2Nb, 2Nd, 2Nf in white) and CTA-South (Confs. 2b, 2c, 2e in green) installations. \textit{(A color version of this figure is available on-line)}.}
	\label{fig:Final_P2_all}
\end{figure}

Here we summarize the results of our simulations for different energy ranges of Conf. 2NN and also for other CTA configurations, such as Confs. 2Nc, 2Ne, 2Nb, 2Nd, 2Nf -- representatives of the northern CTA installation --, and Confs. 2b, 2c, 2e -- possible layouts of CTA-South. The properties of all simulated arrays are listed in Table \ref{tab:a1} and the corresponding values of $\Delta_2$ are shown in Figs. \ref{fig:Final_P2_ArrEn}, \ref{fig:Final_P2_all} and Table \ref{tab:res}. Observations of short-duration with Conf. 2NN ($0.1\times t_{\rm obs}^{\rm M}$, $0.3\times t_{\rm obs}^{\rm M}$, $1\times t_{\rm obs}^{\rm M}$,  $1.5\times t_{\rm obs}^{\rm M}$, $3\times t_{\rm obs}^{\rm M}$) are not sufficient to detect significant pulsations in the 1--10 TeV energy range.

\begin{table}
\caption{Configurations of MAGIC, VERITAS and different sub-arrays of CTA simulated in MC-Prod2.}
\label{tab:a1}
\centering
\begin{tabular}{l c c c c}
\hline\hline
Name 		& Telescopes 		& Energy range 	& $E_{\rm thr}$	& $\left< A_{\rm eff} \right >_{\rm sp}$	\\
			&				& (TeV) 			& (TeV)			& [$10^5 \times \mathrm{m}^2$] \\
\hline
MAGIC 		& $2\times17$ m  	& 0.05--0.4	& 0.072	& 0.07	\\
VERITAS		& $4\times12$ m 	& 0.1--0.4		& 0.136	& 0.28	\\
LST-array		& 4 LST 			& 0.04--158	& 0.040	& 0.49	\\
MST-array	& 14 MST			& 0.1--158	& 0.158	& 0.71	\\
Mini-array	& 9 SST	 		& 1.6--158	& 3.981	& 0.71	\\
&&& \\
Conf. 2NN	& 4 LST 			& 0.04--158	& 0.040	& 0.53	\\
			& 14 MST 		&			& 		&		\\
&&& \\
Conf. 2Nb 	& 4 LST 			& 0.04--100	& 0.040	& 0.53	\\
			& 14 MST 		&			& 		&		\\
&&& \\
Conf. 2Nc 	& 4 LST 			& 0.04--158	& 0.040	& 0.53	\\
			& 10 MST 		&			& 		&		\\
			& 10 SST			&			& 		&		\\
&&& \\
Conf. 2Nd	& 3 LST 			& 0.03--100	& 0.040	& 0.53	\\
			& 12 MST 		&			& 		&		\\
&&& \\
Conf. 2Ne	& 3 LST 			& 0.04--100	& 0.040	& 0.53	\\
			& 12 MST 		&			& 		&		\\
&&& \\
Conf. 2Nf 	& 4 LST 			& 0.04--100	& 0.040	& 0.53	\\
			& 10 MST 		&			& 		&		\\
&&& \\
Conf. 2b		& 3 LST 			& 0.04--158	& 0.040	& 0.24	\\
			& 18 MST			&			& 		&		\\
			& 72 SST			&			& 		&		\\
&&& \\
Conf. 2c		& 3 LST 			& 0.04--158	& 0.040	& 0.26	\\
			& 32 MST			&			& 		&		\\
			& 38 SST			&			& 		&		\\
&&& \\
Conf. 2e		& 4 LST 			& 0.04--158	& 0.040	& 0.50	\\
			& 24 MST			&			& 		&		\\
			& 72 SST			&			& 		&		\\
\hline
\multicolumn{5}{p{0.45\textwidth}}{\textbf{Notes.} Confs. 2NN, 2Nc, 2Ne, 2Nb, 2Nd, 2Nf are representatives of the northern CTA installation. Confs. 2b, 2c and 2e refer to CTA-South. All these arrays are taken from the MC-Prod2 DESY simulation package (\url{http://www.cta-observatory.org/ctawpcwiki/index.php/WP_MC\#Interface_to_WP_PHYS}). LST: Large Size Telescope with diameter 23 m. MST: Medium Size Telescope with diameter 12 m. SST: Small Size Telescope with diameter 4 m. As the best representation for the ASTRI mini-array (Mini-array), we consider a configuration of 9 SST from the same MC-Prod2 simulations (Conf. s9-4-257m). The energy ranges for all these configurations are taken from the corresponding instrument response functions, while those of MAGIC and VERITAS correspond to the energies at which the Crab pulsar spectrum was measured (see \citet{Aleksic2012} and \citet{Aliu2011}, respectively). $E_{\rm thr}$ is the energy threshold, while $\left< A_{\rm eff} \right >_{\rm sp}$ is the spectrum-weighted effective area of each configuration.}
\end{tabular}
\end{table}

\begin{table*}
\caption{Uncertainty in the position of the interpulse P2 ($\Delta_2$ $\mu$s) of the simulated VHE profile of the Crab pulsar, calculated for MAGIC, VERITAS and the different CTA instrumental configurations shown in Figs. \ref{fig:Final_P2_A}, \ref{fig:Final_P2_ArrEn} and \ref{fig:Final_P2_all}. Columns refer to different observing times in units of the MAGIC observing time ($t_{\rm obs}^{\rm M} = 73$ hours). The spectral index of the Crab pulsar spectrum used in the simulations is $\Gamma$=3.57 \citep{Aleksic2012}.}
\label{tab:res}
\centering
\begin{tabular}{l c c c c c c c}
\hline\hline
Name	& $0.1\times t_{\rm obs}^{\rm M}$ & $0.3\times t_{\rm obs}^{\rm M}$ & $1\times t_{\rm obs}^{\rm M}$ & $1.5\times t_{\rm obs}^{\rm M}$ & $3\times t_{\rm obs}^{\rm M}$ & $10\times t_{\rm obs}^{\rm M}$ & $30\times t_{\rm obs}^{\rm M}$ \\
\hline
MAGIC-sim		& $190 \pm 60$	& $110\pm25$	& $60\pm9$		& $50\pm7$		& $35\pm4$		& $19\pm2$		& $11\pm1$	\\
VERITAS-sim		& - 				& $190\pm70$	& $110\pm20$	& $90\pm20$ 		& $62\pm10$		& $34\pm4$		& $20\pm2$	\\
LST-array			& $45\pm6$ 		& $26\pm3$ 		& $14.5\pm1.5$	& $12\pm1$		& $8.3\pm0.9$	& $4.6\pm0.5$	& $2.6\pm0.3$\\
MST-array		& - 			 	& $160\pm50$ 	& $90\pm20$ 		& $75\pm12$		& $53\pm8$ 		& $59\pm3$ 		& $17\pm2$	\\
Conf. 2NN		& $49\pm7$		& $30\pm4$		& $16\pm2$		& $13\pm1$		& $9.1\pm0.9$	& $5.0\pm0.5$	& $2.9\pm0.3$\\
~~~~~ - (0.04--0.1 TeV)	& $38 \pm 4$	& $22\pm2$		& $12\pm1$		& $9.8\pm1.0$	& $7.0\pm0.7$	& $3.8\pm0.4$	& $2.2\pm0.2$\\
~~~~~ - (0.1--1 TeV)	& $160 \pm 50$	& $100\pm20$	& $53\pm8$		& $43\pm6$		& $30\pm4$		& $17\pm2$		& $10\pm1$	\\
~~~~~ - (1--10 TeV)	& -				& - 				&  - 				& - 				&  - 				& - 				& $320\pm250$\\
Conf. 2Nb		& $47\pm7$		& $27\pm3$		& $15\pm2$		& $12\pm1.2$		& $8.6\pm0.9$	& $4.7\pm0.5$	& $2.7\pm0.3$\\
Conf. 2Nc			& $51\pm7$		& $29\pm3$		& $16\pm2$		& $13.0\pm1.4$	& $9.2\pm1.0$	& $5.0\pm0.5$	& $2.9\pm0.3$\\
Conf. 2Nd		& $38\pm5$		& $22\pm2$		& $11.9\pm1.3$	& $9.8\pm1.0$	& $6.9\pm0.7$	& $3.8\pm0.4$	& $2.2\pm0.2$\\
Conf. 2Ne		& $64\pm10$		& $37\pm5$		& $20\pm2$		& $17\pm2$		& $11.9\pm1.2$	& $6.5\pm0.7$	& $3.7\pm0.4$\\
Conf. 2Nf			& $57\pm8$		& $33\pm4$		& $18\pm2$		& $14.7\pm1.7$	& $10.4\pm1.0$	& $5.7\pm0.6$	& $3.3\pm0.3$\\
2b 				& $100\pm20$	& $59\pm9$		& $32\pm4$		& $27\pm3$		& $19\pm2$		& $10.3\pm1.1$	& $5.9\pm0.6$\\
2c 	 			& $110\pm20$	& $65\pm10$		& $36\pm4$		& $29\pm3$		& $20\pm2$		& $11.3\pm1.2$	& $6.5\pm0.7$\\
2e 				& $58\pm8$		& $33\pm4$		& $18\pm2$		& $15.0\pm1.6$	& $11\pm1$		& $5.8\pm0.6$	& $3.3\pm0.3$\\

\hline
\multicolumn{8}{p{0.72\textwidth}}{\textbf{Notes.} Error-bars represent the standard deviation calculated from a set of simulations. The results for the ASTRI mini-array are not shown because the time required for a significant detection is more than $30\times t_{\rm obs}^{\rm M}$ (see text for details).
}
\end{tabular}
\end{table*}

\bsp	
\label{lastpage}
\end{document}